\documentclass[11pt,preprint,showpacs,preprintnumbers,amsmath,amssymb]{revtex4}

\usepackage{exscale}
\usepackage{relsize}
\usepackage{graphicx}
\usepackage{dcolumn}
\usepackage{bm}
\usepackage{multirow}
\usepackage{CJK}
\usepackage{diagbox}
\usepackage{subfigure}

\begin{document}

\begin{CJK*}{GBK}{}

\title{Study of $CP$ Violations in $B^-\rightarrow K^- \pi^+\pi^-$ and $B^-\rightarrow K^- f_0(500)$ decays in the QCD factorization approach}
\author{Jing-Juan Qi \footnote{e-mail: jjqi@mail.bnu.edu.cn}}
\affiliation{\small{College of Nuclear Science and Technology, Beijing Normal University, Beijing 100875, China}}

\author{Zhen-Yang Wang}
\affiliation{\small{College of Nuclear Science and Technology, Beijing Normal University, Beijing 100875, China}}

\author{Zhen-Hua Zhang \footnote{Corresponding author, e-mail: zhangzh@usc.edu.cn}}
\affiliation{\small{School of Nuclear and Technology, University of South China, Hengyang, Hunan 421001, China}}

\author{Xin-Heng Guo \footnote{Corresponding author, e-mail: xhguo@bnu.edu.cn}}
\affiliation{\small{College of Nuclear Science and Technology, Beijing Normal University, Beijing 100875, China}}

\author{Jing Xu}
\affiliation{\small{Department of Physics, Yantai University, Yantai 264005, China}}

\date{\today\\}

\begin{abstract}
Within the QCD factorization approach, we study the $CP$ violations in $B^-\rightarrow K^-\pi^+\pi^-$ and $B^-\rightarrow K^- f_0(500)$ decays. We find the experimental data of the localized $CP$ asymmetry in $B^-\rightarrow K^-\pi^+\pi^-$ decays in the region $m_{K^-\pi^+}^2<15$ $\mathrm{GeV}^2$ and $0.08<m_{\pi^+\pi^-}^2<0.66$ $\mathrm{GeV}^2$ can be explained by the interference of two intermediate resonances, $\rho^0(770)$ and $f_0(500)$ when the parameters in our interference model are in the allowed ranges, i.e. the relative strong phase $\delta\in[0, 1.745]\cup[3.578, 6.266]$ and the end-point divergence parameters $\rho_S\in[2.790, 5.290]$ and $\phi_S \in [1.518, 5.183]$. With the obtained allowed ranges for $\rho_S$ and $\phi_S$, we obtain the predictions for the $CP$ asymmetry parameter $A_{CP} \in [-0.259, 0.006]$ and the branching fraction $\mathcal{B} \in [0.585, 3.230]\times10^{-5}$ for $B^-\rightarrow K^-f_0(500)$ decay modes.
\end{abstract}

\maketitle
\end{CJK*}

\section{Introduction}
Charge-Parity ($CP$) violation is essential to our understanding of both particle physics and the evolution of the early universe. It is one of the most fundamental and important properties of weak interaction, and has gained extensive attentions ever since its first discovery in 1964 \cite{Christenson:1964fg}. In the Standard Model (SM), $CP$ violation is related to the weak complex phase in the Cabibbo-Kobayashi-Maskawa (CKM) matrix, which describes the mixing of different generations of quarks \cite{Cabibbo:1963yz, Kobayashi:1973fv}. Besides the weak phase, a large strong phase is also needed for a large \emph{CP} asymmetry. Generally, this strong phase is provided by QCD loop corrections and some phenomenological models.

In recent years, prompted by a large number of experimental measurements, three-body hadronic $B$ meson decays have been studied by using different theoretical frameworks \cite{Gronau:2003ep,Krankl:2015fha,Calderon:2015lsg, Wang:2016rlo, Nogueira:2016qwj}. Strong dynamics contained in three-body hadronic $B$ meson decays is much more complicated than that in two-body cases, e.g. how to factorize $B$ to three-body final states matrix elements. Both $BABAR$ \cite{Aubert:2008bj} and Belle \cite{Garmash:2005rv} Collaborations claimed evidence of partial rate asymmetries in the channels $B^{\pm}\rightarrow \rho^0(770)K^{\pm}$ in the Dalitz plot analysis of $B^-\rightarrow K^-\pi^+\pi^-$. LHCb also observed the large $CP$ asymmetry in the localized region of the phase space \cite{Aaij:2013sfa, Aaij:2013bla}, $A_\emph{CP}(K^-\pi^+\pi^-)=0.678\pm0.078\pm0.0323\pm0.007$, for $m_{K^-\pi^+}^2<15$ $\mathrm{GeV}^2$ and $0.08<m_{\pi^+\pi^-}^2<0.66$ $\mathrm{GeV}^2$, which spans the $\pi^+\pi^-$ masses around the $\rho^0(770)$ resonance. Such three-body decays in this region have been studied in Refs. \cite{Cheng:2013dua,Cheng:2016ajl} using a simple model based on the framework of the factorization approach. The authors of Refs \cite{Bhattacharya:2013boa,Zhang:2013oqa} considered the possibility of obtaining a large local $CP$ violation in $B^-\rightarrow\pi^+\pi^-\pi^-$ decay from the interference of the resonances $\rho^0(770)$ and $f_0(500)$. In this work, we will apply this mechanism to study $CP$ violation in $B^-\rightarrow K^-\pi^+\pi^-$ decay with the interference of $\rho^0(770)$ and $f_0(500)$ in the region of $m_{K^-\pi^+}^2<15$ $\mathrm{GeV}^2$ and $0.08<m_{\pi^+\pi^-}^2<0.66$ $\mathrm{GeV}^2$.

In contrast to vector and tensor mesons, the identification of scalar mesons is a long-standing puzzle, because some of them have large decay widths which cause strong overlaps between resonances and backgrounds in experiments \cite{Agashe:2014kda}. Up to now, there have been some progresses in the study of charmless hadronic $B$ decays with scalar mesons in the final states both experimentally and theoretically. On the experimental side, measurements of $B$ decays to the scalar mesons such as $f_0(980)$, $f_0(1370)$, $f_0(1500)$, $f_0(1710)$, $a_0(980)$, $a_0(1450)$, and $K_0^*(1430)$ have been reported by $BABAR$ and Belle Collaborations, but the decays to $f_0(500)$ have not been reported and the $CP$ violation and the branching fractions have not been measured for such processes. So it is important to predict the values of $A_{CP}(B^-\rightarrow K^-f_0(500))$ and $\mathcal{B}(B^-\rightarrow K^-f_0(500))$. Although the light scalar mesons are widely perceived as primarily the 4-quark bound states, in practice it is difficult to make quantitative predictions based on the 4-quark picture for the light scalar mesons, hence, predictions are made in the 2-quark model for the $f_0(500)$ meson \cite{Cheng:2005nb}.

Theoretically, to calculate the hadronic matrix elements of $B$ nonleptonic weak decays, some approaches, including the naive factorization \cite{Wirbel:1985ji,Bauer:1986bm}, the QCD factorization (QCDF) \cite{Beneke:2006hg,Beneke:1999br,Cheng:2008gxa}, the perturbative QCD (PQCD) approach \cite{Keum:2000ph,Keum:2000wi,Lu:2000em}, and the soft-collinear effective theory (SCET) \cite{Bauer:2001cu,Bauer:2001yt}, have been fully developed and extensively employed in recent years. In this work, within the framework of QCDF \cite{Cheng:2013fba,Li:2015zra}, we will study the decays of  $B^-\rightarrow K^-\pi^+\pi^-$ via the interference of $\rho^0(770)$ and $f_0(500)$ and $B^-\rightarrow K^-f_0(500)$.

The remainder of this paper is organized as follows. In Sect. ${\mathrm{\uppercase\expandafter{\romannumeral2}}}$, we briefly present the formalism of the QCD factorization approach. In Sect. ${\mathrm{\uppercase\expandafter{\romannumeral3}}}$, we present the formalisms for CP violation of $B^-\rightarrow K^-f_0(500)$ and $B^-\rightarrow K^-\pi^+\pi^-$. The numerical results are given in Sect. ${\mathrm{\uppercase\expandafter{\romannumeral4}}}$ and we summarize our work in Sect ${\mathrm{\uppercase\expandafter{\romannumeral5}}}$.

\section{QCD FACTORIZATION}
With the operator product expansion, the effective weak Hamiltonian for $B$ meson decays can be written as \cite{Beneke:2003zv}
 \begin{equation}\label{Hamiltonian}
 \mathcal{H}_{eff}=\frac{G_F}{\sqrt{2}}\sum_{p=u,c}\sum_{D=d,s}\lambda_{p}^{(D)}{(C_1Q_1^p+C_2Q_2^p+\sum_{i=3}^{10}C_iQ_i+C_{7\gamma}Q_{7\gamma}+C_{8g}Q_{8g})+h.c.},
 \end{equation}
 where $G_F$ represents the Fermi constant, $\lambda_p^{(D)}=V_{pb}V_{pD}^*$, $V_{pb}$ and $V_{pD}$ are the CKM matrix elements, $C_i$ $(i=1,2,\cdots,10)$ are the Wilson coefficients, $Q_{1,2}^p$ are the tree level operators and $Q_{3-10}$ are the penguin ones, and $Q_{7\gamma}$ and $Q_{8g}$ are the electromagnetic and chromomagnetic dipole operators, respectively. The explicit forms of the operators $Q_i$ are \cite{Beneke:2001ev}
\begin{equation}\label{qi}
\begin{split}
&Q_1^p=\bar{p}\gamma_\mu(1-\gamma_5)b\bar{D}\gamma^\mu(1-\gamma_5)p, \quad\quad\quad\quad\quad Q_2^p=\bar{p}_\alpha\gamma_\mu(1-\gamma_5)b_\beta\bar{D}_\beta\gamma^\mu(1-\gamma_5)p_\alpha, \\
&Q_3=\bar{D}\gamma_\mu(1-\gamma_5)b\sum_{q'}\bar{q'}\gamma^\mu(1-\gamma_5)q',\quad\quad\quad Q_4=\bar{D}_\alpha\gamma_\mu(1-\gamma_5)b_\beta\sum_{q'}\bar{q'}_\beta\gamma^\mu(1-\gamma_5)q'_\alpha, \\
&Q_5=\bar{D}\gamma_\mu(1-\gamma_5)b\sum_{q'}\bar{q'}\gamma^\mu(1+\gamma_5)q',\quad\quad\quad Q_6=\bar{D}_\alpha\gamma_\mu(1-\gamma_5)b_\beta\sum_{q'}\bar{q'}_\beta\gamma^\mu(1+\gamma_5)q'_\alpha,\\
&Q_7=\frac{3}{2}\bar{D}\gamma_\mu(1-\gamma_5)b\sum_{q'}e_{q'}\bar{q'}\gamma^\mu(1+\gamma_5)q',\quad\, Q_8=\frac{3}{2}\bar{D}_\alpha\gamma_\mu(1-\gamma_5)b_\beta\sum_{q'}e_{q'}\bar{q'}_\beta\gamma^\mu(1+\gamma_5)q'_\alpha,\\
&Q_9=\frac{3}{2}\bar{D}\gamma_\mu(1-\gamma_5)b\sum_{q'}e_{q'}\bar{q'}\gamma^\mu(1-\gamma_5)q',\quad\, Q_{10}=\frac{3}{2}\bar{D}_\alpha\gamma_\mu(1-\gamma_5)b_\beta\sum_{q'}e_{q'}\bar{q'}_\beta\gamma^\mu(1-\gamma_5)q'_\alpha,\\
&Q_{7\gamma}=\frac{-e}{8\pi^2}m_b\bar{s}\sigma_{\mu\nu}(1+\gamma_5)F^{\mu\nu}b,\quad \quad\quad\quad\quad\quad Q_{8g}=\frac{-g_s}{8\pi^2}m_b\bar{s}\sigma_{\mu\nu}(1+\gamma_5)G^{\mu\nu}b,\\
\end{split}
 \end{equation}
where $\alpha$ and $\beta$ are color indices, $q'=u,d,s,c$ or $b$ quarks.

In dealing with the charmless $B$ decay into two mesons $M_1$ and $M_2$, the decay amplitude is usually divided into the emission part and the annihilation part in terms of the structures of the topological diagrams. In the heavy quark limit, the former part can be written as the product of the decay constant and the form factor, while for the latter part, it is always regarded as being power suppressed. With the standard procedure of the QCDF, the emission part of the decay amplitude has the following form:
\begin{equation}\label{M1M2}
\mathcal{M}(B^-\rightarrow M_1M_2)=\frac{G_F}{\sqrt{2}}\sum_{p=u,c}\sum_{i}V_{pb}V_{pd}^*\alpha^p_i(\mu)\times\langle{M_1M_2}|Q_i|B\rangle,
\end{equation}
where $\alpha^p_i(\mu)$ are flavour parameters which can be expressed in terms of the effective parameters $a_i^p$, which can be calculated perturbatively, with the expressions given by \cite{Beneke:2003zv}
\begin{equation}\label{a}
\begin{split}
a_i^p{(M_1M_2)}&={(C'_i+\frac{C'_{i\pm1}}{N_c})}N_i{(M_2)}+\frac{C'_{i\pm1}}{N_c}\frac{C_F\alpha_s}{4\pi}{\bigg[V_i{(M_2)}+\frac{4\pi^2}{N_c}H_i{(M_1M_2)}\bigg]+P_i^p{(M_2)}},
\end{split}
 \end{equation}
 where $C'_i$ are effective Wilson coefficients which are defined as $C_i(m_b)\langle Q_i(m_b)\rangle=C'_i\langle Q_i\rangle^{tree}$ with $\langle Q_i\rangle^{tree}$ being the matrix element at the tree level, the upper (lower) signs apply when $i$ is odd (even), $N_i{(M_2)}$ are leading-order coefficients, $C_F={(N_c^2-1)}/{2N_c}$ with $N_c=3$, the quantities $V_i{(M_2)}$ account for one-loop vertex corrections, $H_i{(M_1M_2)}$ describe hard spectator interactions with a hard gluon exchange between the emitted meson and the spectator quark of the $B$ meson, and $P_i^p{(M_1M_2)}$ are from penguin contractions \cite{Beneke:2003zv}. Similarly, weak annihilation contributions are described by the terms $b_i$ and $b_{i,EW}$, which have the following expressions:
\begin{equation}\label{b}
\begin{split}
&b_1=\frac{C_F}{N_c^2}C'_1A_1^i, \quad b_2=\frac{C_F}{N_c^2}C'_2A_1^i, \\
&b_3^p=\frac{C_F}{N_c^2}\bigg[C'_3A_1^i+C'_5(A_3^i+A_3^f)+N_cC'_6A_3^f \bigg],\quad b_4^p=\frac{C_F}{N_c^2}\bigg[C'_4A_1^i+C'_6A_2^i \bigg], \\
&b_{3,EW}^p=\frac{C_F}{N_c^2}\bigg[C'_9A_1^i+C'_7(A_3^i+A_3^f)+N_cC'_8A_3^f \bigg],\\
&b_{4,EW}^p=\frac{C_F}{N_c^2}\bigg[C'_{10}A_1^i+C'_8A_2^i \bigg],
\end{split}
 \end{equation}
 where the subscripts 1, 2, 3 of $A_n^{i,f}(n=1,2,3)$ stand for the annihilation amplitudes induced from $(V-A)(V-A)$, $(V-A)(V+A)$, and $(S-P)(S+P)$ operators, respectively, and the superscripts $i$ and $f$ refer to gluon emission from the initial- and final-state quarks, respectively. The explicit expressions for $A_n^{i,f}$ can be found in Ref. \cite{Beneke:2003zv}.
When dealing with the weak annihilation contributions and the hard spectator contributions, one suffers from the infrared endpoint singularity $X=\int_0^1 dx/(1-x)$. The treatment of the endpoint divergence is model dependent, and we follow Ref.  \cite{Beneke:1999br} to parameterize the endpoint divergence in the annihilation diagrams as
\begin{equation}\label{X}
\begin{split}
X=(1+\rho e^{i\phi})\ln\frac{m_B}{\Lambda_h},
 \end{split}
 \end{equation}
where $\Lambda_h$ is a typical scale of order 500 MeV, $\rho$ is an unknown real parameter, $\phi$ is the free strong phase in the range $[0,2\pi]$. The QCDF approach itself cannot give information or constraints on the phenomenological parameters $\rho$ and $\phi$, both of them should be fixed by experimental data such as branching fractions and CP asymmetries.

\section{CALCULATION OF CP VIOLATION}

\subsection{$CP$ violation formalism for $B^-\rightarrow K^-\pi^+\pi^-$}
In this section, we will consider a $B$ meson three-body decay process, $B\rightarrow M_1M_2M_3$, where $M_i$ $(i=1, 2, 3)$ are light mesons. There are two resonaces, $X$ and $Y$, appearing during this process: $B\rightarrow X(Y)M_3$, then both $X$ and $Y$ decay to $M_1M_2$. The amplitude for $B\rightarrow X(Y)M_3\rightarrow M_1M_2M_3$ around the $X$ and $Y$ resonance region can be expressed as \cite{Zhang:2013oqa}
\begin{equation}
\mathcal{M}=\mathcal{M}_X+\mathcal{M}_Ye^{i\delta},
 \end{equation}
 where $\delta$ is a relative strong phase, $\mathcal{M}_X$ and $\mathcal{M}_Y$ are the amplitudes for $B\rightarrow XM_3\rightarrow M_1M_2M_3$ and $B\rightarrow YM_3\rightarrow M_1M_2M_3$, respectively, and they take the following form:
 \begin{equation} \label{Hstrong}
\begin{split}
\mathcal{M}_X=\frac{\langle X M_3|\mathcal{H}_{eff}|B\rangle \langle M_1M_2|\mathcal{H}_{XM_1M_2}|X\rangle}{S_{X}},\\
\mathcal{M}_Y=\frac{\langle Y M_3|\mathcal{H}_{eff}|B\rangle \langle M_1M_2|\mathcal{H}_{YM_1M_2}|Y\rangle}{S_{Y}}. \\
 \end{split}
 \end{equation}
In the above equations, $\mathcal{H}_{X(Y)M_1M_2}$ is the strong Hamiltonian for the transition $X(Y)\rightarrow M_1M_2$, $S_{X(Y)}$ is the reciprocal of the propagator of $X(Y)$ which takes the form $s_{12}-m_{X(Y)}^2+i\sqrt{s_{12}}\Gamma_{X(Y)}$, where $s_{ij}$ $(i,j=1,2,3)$ is the invariant mass squared of mesons $M_i$ and $M_j$ \cite{Zhang:2013oqa}.
For the specific process $B^-\rightarrow K^-\pi^+\pi^-$ in the region $m_{K^\pm\pi^\pm}^2<15$ $\mathrm{GeV}^2$ and $0.08<m_{\pi^+\pi^-}^2<0.66$ $\mathrm{GeV}^2$, $\rho^0(770)$ and $f_0(500)$ are the dominate resonances. The effective Hamiltonians for strong processes $\rho^0(770)\rightarrow \pi^+\pi^-$ and $f_0(500)\rightarrow \pi^+\pi^-$ can be formally expressed as
\begin{equation}\label{Hs}
\begin{split}
\mathcal{H}_{\rho^0\pi\pi}&=-ig_{\rho^0\pi\pi}\rho_\mu^0\pi^+ \overleftrightarrow {\partial}^\mu \pi^-,\\
\mathcal{H}_{f_0\pi\pi}&=g_{f_0\pi\pi}f_0(2\pi^+\pi^-+\pi^0\pi^0),
 \end{split}
 \end{equation}
where $\rho_\mu^0$, $f_0$ and $\pi^\pm$ are the field operators for $\rho^0(770)$, $f_0(500)$ and $\pi$ mesons, respectively, $g_{\rho^0\pi\pi}$ and $g_{f_0\pi\pi}$ are the effective coupling constants which can be expressed in terms of the decay widths of $\rho^0\rightarrow\pi^+\pi^-$ and $f_0\rightarrow\pi^+\pi^-$, respectively,
\begin{equation}
\begin{split}
g_{\rho^0\pi\pi}^2&=\frac{48\pi}{(1-\frac{4m_\pi^2}{m_\rho^2})^{3/2}}\times\frac{\Gamma_{\rho^0\rightarrow\pi^+\pi^-}}{m_\rho},\\
g_{f_0\pi\pi}^2&=\frac{4\pi m_{f_0}\Gamma_{f_0\rightarrow\pi^+\pi^-}}{(1-\frac{4m_\pi^2}{m_{f_0}^2})^{1/2}}.
 \end{split}
 \end{equation}
Both $\rho^0(770)$ and $f_0(500)$ decay into one pion pair dominantly through the strong interaction, and the isospin symmetry of the strong interaction tells us that  $\Gamma_{\rho^0}\approx\Gamma_{\rho^0\rightarrow\pi^+\pi^-}$ and $\Gamma_{f_0}\approx\frac{3}{2}\Gamma_{f_0\rightarrow\pi^+\pi^-}$.

The differential $CP$ asymmetry parameter can be defined as
 \begin{equation}\label{CP asymmetry parameter}
A_{CP}=\frac{|\mathcal{M}|^2-|\mathcal{\bar{M}}|^2}{|\mathcal{M}|^2+|\mathcal{\bar{M}}|^2}.
 \end{equation}
 By integrating the denominator and numerator of $A_{CP}$ in the region $R$, we get the localized integrated $CP$ asymmetry, which can be measured in experiments and takes the following form:
  \begin{equation}\label{localized CP}
A_{CP}^R=\frac{\int_R ds_{12}ds_{13}(|\mathcal{M}|^2-|\mathcal{\bar{M}}|^2)}{\int_R ds_{12}ds_{13}(|\mathcal{M}|^2+|\mathcal{\bar{M}}|^2)},
 \end{equation}
 where $R$ represents certain region of the phase space, in our work $R$ includes $m_{K^-\pi^+}^2<15$ $\mathrm{GeV}^2$ and $0.08<m_{\pi^+\pi^-}^2<0.66$ $\mathrm{GeV}^2$ in $B^-\rightarrow K^-\pi^+\pi^-$ decay.
\subsection{Calculation of amplitudes of $B^-\rightarrow \rho^0(770)(f_0(500))K^-\rightarrow\pi^+\pi^-K^-$}
 In the QCDF, including the emission and annihilation contributions, the decay amplitudes of $B^-\rightarrow\rho^0(770) K^-$ and $B^-\rightarrow f_0(500) K^-$ can be finally given as
\begin{small}
\begin{equation}
\begin{split}
\mathcal{M}(B^-\rightarrow\rho^0(770) K^-)&=\langle \rho^0(770) K^-|\mathcal{H}_{eff}|B^-\rangle\\
&=\sum_{p=u,c}\lambda_p^{(s)}\frac{-iG_F}{2}\bigg\{\Big[\alpha_1(\rho K)\delta_{pu}+\alpha_4^p(\rho K)+\alpha_{4,EW}^p(\rho K)\Big]\\
&\times m_B^2A_0^{B\rightarrow\rho}(0)f_K+\Big[\alpha_2(K \rho)\delta_{pu}+\frac{3}{2}\alpha_{3,EW}^p(K \rho)\Big]\times m_B^2F_+^{B\rightarrow f}(0)f_\rho\\
&+\Big[b_2(\rho K)\delta_{pu}+b_3^p(\rho K)+b_{3,EW}^p(\rho K)\Big]\times f_Bf_\rho f_K\bigg\},\\
\end{split}
\end{equation}
\end{small}
for $B^-\rightarrow\rho^0(770) K^-$, and
\begin{small}
\begin{equation}\label{Mbfk}
\begin{split}
\mathcal{M}(B^-\rightarrow f_0(500) K^-)&=\langle f_0(500) K^-|\mathcal{H}_{eff}|B^-\rangle\\
&=\sum_{p=u,c}\lambda_p^{(s)}\frac{G_F}{2}\bigg\{\left[\alpha_1(f K)\delta_{pu}+\alpha_4^p(f K)+
\alpha_{4,EW}^p(f K)\right]\times(m_{f}^2-m_B^2)F_0^{B\rightarrow f}(m_K^2)f_K\\
&+\Big[\alpha_2(K f)\delta_{pu}+2\alpha_{3}(K f)+\frac{1}{2}\alpha_{3,EW}^p(K f))\Big]\times (m_B^2-m_{K}^2)F_0^{B\rightarrow K}(m_{f}^2)\bar{f}^u_{f_0(500)}\\
&+\Big[\sqrt{2}\alpha_3^p(K f)+\sqrt{2}\alpha_{4}^p(K f)-\frac{1}{\sqrt{2}}\alpha_{3,EW}^p(K f)-\frac{1}{\sqrt{2}}\alpha_{4,EW}^p(K f)\Big]\\
&\times (m_B^2-m_{K}^2)F_0^{B\rightarrow K}(m_{f}^2)\bar{f}^s_{f_0(500)}-\Big[b_2(f K)\delta_{pu}+b_3^p(f K)+b_{3,EW}^p(f K)\Big]\\
&\times f_Bf_K\bar{f}^u_{f_0(500)}-\sqrt{2}\Big[b_2(K f)\delta_{pu}+b_3^p(K f )+b_{3,EW}^p(K f)\Big]\times f_Bf_K\bar{f}^s_{f_0(500)}\bigg\},\\
\end{split}
\end{equation}
\end{small}
for $B^-\rightarrow f_0(500) K^-$, where $\rho$ and $f$ are the abbreviations for $\rho^0(770)$ and $f_0(500)$, respectively, $A_0^{B\rightarrow M_1}(0)$ and $F_{+,0}^{B\rightarrow M_2}(q^2)$ are form factors for $B$ to $M_1$ and $M_2$ transitions, $f_K$, $f_\rho$ and $f_B$ are decay constants of $K$, $\rho$ and $B$ mesons, respectively, $\bar{f}^u_{f_0(500)}$ and $\bar{f}^s_{f_0(500)}$ are decay constants of $f_0(500)$ coming from the up and strange quark components, respectively.

From Eq. (\ref{Hs}), we can obtain the amplitudes for $\rho^0(770)\rightarrow\pi^+\pi^-$ and $f_0(500)\rightarrow\pi\pi$ as
\begin{equation}\label{Mstrong}
\begin{split}
\mathcal{M}(\rho^0(770)\rightarrow\pi^+\pi^-)&=\langle \pi^+\pi^-|\mathcal{H}_{\rho^0\pi\pi}|\rho^0(770)\rangle=g_{\rho\pi^+\pi^-}\varepsilon_{\rho^0}\cdot(p_{\pi^-}-p_{\pi^+}),\\
  \mathcal{M}(f_0(500)\rightarrow\pi^+\pi^-)&=\langle \pi^+\pi^-|\mathcal{H}_{f_0\pi\pi}|f_0(500)\rangle=2g_{f_0\pi^+\pi^-},\\
\end{split}
\end{equation}
where $\varepsilon_{\rho^0}$ is the polarization vector of $\rho^0(770)$, $p_{\pi^+}$ and $p_{\pi^-}$ are the momenta of $\pi^+$ and $\pi^-$, respectively.

Considering the total processes, one can get
\begin{equation}\label{M1}
\begin{split}
  &\mathcal{M}(B^-\rightarrow\rho^0(770) K^-\rightarrow\pi^+\pi^- K^-)
  =\sum_{p=u,c}\lambda_p^{(s)}\frac{-iG_Fg_{\rho\pi^+\pi^-}}{S_{\rho^0(770)}}(\hat{s}_{K\pi}-s_{K\pi})\\
  &\times\bigg\{A_0^{B\rightarrow\rho}(0)f_K\left[\alpha_1(\rho K)\delta_{pu}+\alpha_4^p(\rho K)+\alpha_{4,EW}^p(\rho K)\right]+F_+^{B\rightarrow K}(0)f_\rho \left[\alpha_2(K \rho)\delta_{pu}+\frac{3}{2}\alpha_{3,EW}^p(K \rho)\right]\\
  &+\frac{f_B f_\rho f_K}{m_B^2} \left[b_2(\rho K)\delta_{pu}+b_3^p(\rho K)+b_{3,EW}^p(\rho K)\right]\bigg\},\\
\end{split}
\end{equation}
for the $B^-\rightarrow\rho^0(770) K^-\rightarrow\pi^+\pi^- K^-$ decay mode, where $\hat{s}_{K\pi}$ is the midpoint of the allowed range of $s_{K^-\pi^+}$, i.e. $\hat{s}_{K\pi}=(s_{K^-\pi^+, \textrm{max}}+s_{K^-\pi^+,\textrm{min}})/2$, with $s_{K^-\pi^+,\textrm{max}}$ and $s_{K^-\pi^+,\textrm{min}}$ being the maximum and minimum values of $s_{K^-\pi^+}$ for fixed $s_{\pi+\pi^-}$.

For the $B^-\rightarrow f_0(500) K^-\rightarrow\pi^+\pi^- K^-$ decay modes.
\begin{small}
\begin{equation}\label{M2}
\begin{split}
  &\mathcal{M}(B^-\rightarrow f_0(500) K^-\rightarrow\pi^+\pi^- K^-)=\sum_{p=u,c}\lambda_p^{(s)}\frac{G_Fg_{f\pi^+\pi^-}}{S_{f_0(500)}}
  \bigg\{(m_f^2-m_B^2)F_0^{B\rightarrow f}(m_K^2)\\
  &\times f_K\left[\delta_{pu}\alpha_1(fK)+\alpha_4^p(fK)+\alpha_{4,EW}^p(fK)\right]-f_Bf_K\bar{f}^u_{f_0(500)}\left[\delta_{pu}b_2(fK)+b_3^p(fK)+b_{3,EW}^p(fK)\right]\\
  & +(m_B^2-m_K^2)F_0^{B\rightarrow K}(m_f^2)\bar{f}^u_{f_0(500)}\left[\delta_{pu}\alpha_2(K f)+2\alpha_3^p(Kf)+\frac{1}{2}\alpha_{3,EW}^p(Kf)\right]\\
  &+(m_B^2-m_K^2)F_0^{B\rightarrow K}(m_f^2)\bar{f}^s_{f_0(500)}\left[\sqrt{2}\alpha_3^p(K f)+\sqrt{2}\alpha_4^p(K f)-\frac{1}{\sqrt{2}}\alpha_{3,EW}^p(K f)-\frac{1}{\sqrt{2}}\alpha_{4,EW}^p(K f)\right]\\
  &-f_Bf_K\bar{f}^s_{f_0(500)}\left[\sqrt{2}\delta_{pu}b_2(Kf)+\sqrt{2}b_3^p(Kf)+\sqrt{2}b_{3,EW}^p(K f)\right]\bigg\}.\\
\end{split}
\end{equation}
\end{small}

The amplitude for $B^-\rightarrow K^-\pi^+\pi^-$ around the $f_0(500)$ and $\rho^0(770)$ resonance region can be expressed as
\begin{small}
\begin{equation}\label{M}
\begin{split}
 \mathcal{M}&=\mathcal{M}(B^-\rightarrow f_0(500) K^-\rightarrow\pi^+\pi^- K^-)+\mathcal{M}(B^-\rightarrow\rho^0(770) K^-\rightarrow\pi^+\pi^- K^-)e^{i\delta}\\
 &=\sum_{p=u,c}\lambda_p^{(s)}\frac{G_Fg_{f\pi^+\pi^-}}{S_{f_0(500)}}
  \bigg\{(m_f^2-m_B^2)F_0^{B\rightarrow f}(0)f_K\left[\delta_{pu}\alpha_1(fK)+\alpha_4^p(fK)+\alpha_{4,EW}^p(fK)\right]\\
  &-f_Bf_K\bar{f}^u_{f_0(500)}\left[\delta_{pu}b_2(fK)+b_3^p(fK)+b_{3,EW}^p(fK)\right]\\
  &-f_Bf_K\bar{f}^s_{f_0(500)}\left[\sqrt{2}\delta_{pu}b_2(Kf)+\sqrt{2}b_3^p(Kf)+\sqrt{2}b_{3,EW}^p(K f)\right]\\
  &+(m_B^2-m_K^2)F_0^{B\rightarrow K}(0)\bar{f}^u_{f_0(500)}\left[\delta_{pu}\alpha_2(K f)+2\alpha_3^p(Kf)+\frac{1}{2}\alpha_{3,EW}^p(Kf)\right]\\
  &+(m_B^2-m_K^2)F_0^{B\rightarrow K}(0)\bar{f}^s_{f_0(500)}\left[\sqrt{2}\alpha_3^p(K f)+\sqrt{2}\alpha_4^p(K f)-\frac{1}{\sqrt{2}}\alpha_{3,EW}^p(K f)-\frac{1}{\sqrt{2}}\alpha_{4,EW}^p(K f)\right]\bigg\}\\
  &+\sum_{p=u,c}\lambda_p^{(s)}\frac{-iG_Fg_{\rho\pi^+\pi^-}}{S_{\rho^0(770)}}(\hat{s}_{K\pi}-s_{K\pi})\bigg\{A_0^{B\rightarrow\rho}(0)f_K\left[\alpha_1(\rho K)\delta_{pu}+\alpha_4^p(\rho K)+\alpha_{4,EW}^p(\rho K)\right]\\
  &+F_0^{B\rightarrow K}(0)f_\rho \left[\alpha_2(K \rho)\delta_{pu}+\frac{3}{2}\alpha_{3,EW}^p(K \rho)\right]+\frac{f_B f_\rho f_K}{m_B^2} \left[b_2(\rho K)\delta_{pu}+b_3^p(\rho K)+b_{3,EW}^p(\rho K)\right]\bigg\}e^{i\delta},\\
\end{split}
\end{equation}
\end{small}
where $\delta\in[0,2\pi]$.
Substituting Eq. (\ref{M}) into Eq. (\ref{localized CP}) and taking the integral region $R$ as $m_{K^-\pi^+}^2<15$ $\mathrm{GeV}^2$ and $0.08<m_{\pi^+\pi^-}^2<0.66$ $\mathrm{GeV}^2$, we can get the expression of the localized $A_{CP}(B\rightarrow K^- \pi^+\pi^-)$, which is a function of $X (\rho_S,\phi_S)$ and $\delta$.
\subsection{Calculation of differential $CP$ violation and branching fraction of $B^-\rightarrow K^- f_0(500)$}

Using Eq. (\ref{CP asymmetry parameter}), the differential $CP$ asymmetry parameter of $B\rightarrow M_1M_2$ can be expressed as
 \begin{equation}\label{CP asymmetry}
A_{CP}(B\rightarrow M_1M_2)=\frac{|\mathcal{M}(B\rightarrow M_1M_2)|^2-|\mathcal{\bar{M}}(B\rightarrow M_1M_2)|^2}{|\mathcal{M}(B\rightarrow M_1M_2)|^2+|\mathcal{\bar{M}}(B\rightarrow M_1M_2)|^2}.
 \end{equation}
The branching fraction of $B\rightarrow M_1M_2$ decay has the following form:
\begin{equation}\label{e1}
\mathcal{B}(B\rightarrow M_1M_2)=\tau_B \frac{|p_c|}{8\pi m_B^2}|\mathcal{M}(B\rightarrow M_1M_2)|^2,
\end{equation}
where $\tau_B$ is the lifetime of $B$ meson, $m_B$ is the mass of $B$ meson, $|p_c|$ is the norm of a hadron's three momentum in the final state which can be expressed as
\begin{equation}\label{e2}
|p_c|=\frac{1}{2m_B}\sqrt{[m_B^2-(m_{M_1}+m_{M_2})^2][m_B^2-(m_{M_1}-m_{M_2})^2]},
\end{equation}
where $m_{M_1}$ and $m_{M_2}$ are the two final state mesons' masses, respectively.

Substituting the amplitude of $B^-\rightarrow K^- f_0(500)$,
\begin{small}
\begin{equation}\label{Mbfk1}
\begin{split}
\mathcal{M}(B^-\rightarrow f_0(500) K^-)&=\langle f_0(500) K^-|\mathcal{H}_{eff}|B^-\rangle\\
&=\sum_{p=u,c}\lambda_p^{(s)}\frac{G_F}{2}\bigg\{\left[\alpha_1(f K)\delta_{pu}+\alpha_4^p(f K)+
\alpha_{4,EW}^p(f K)\right]\times(m_{f}^2-m_B^2)F_0^{B\rightarrow f}(m_K^2)f_K\\
&+\Big[\alpha_2(K f)\delta_{pu}+2\alpha_{3}(K f)+\frac{1}{2}\alpha_{3,EW}^p(K f))\Big]\times (m_B^2-m_{K}^2)F_0^{B\rightarrow K}(m_{f}^2)\bar{f}^u_{f_0(500)}\\
&+\Big[\sqrt{2}\alpha_3^p(K f)+\sqrt{2}\alpha_{4}^p(K f)-\frac{1}{\sqrt{2}}\alpha_{3,EW}^p(K f)-\frac{1}{\sqrt{2}}\alpha_{4,EW}^p(K f)\Big]\\
&\times (m_B^2-m_{K}^2)F_0^{B\rightarrow K}(m_{f}^2)\bar{f}^s_{f_0(500)}-\Big[b_2(f K)\delta_{pu}+b_3^p(f K)+b_{3,EW}^p(f K)\Big]\\
&\times f_Bf_K\bar{f}^u_{f_0(500)}-\sqrt{2}\Big[b_2(K f)\delta_{pu}+b_3^p(K f )+b_{3,EW}^p(K f)\Big]\times f_Bf_K\bar{f}^s_{f_0(500)}\bigg\},\nonumber
\end{split}
\end{equation}
\end{small}
into Eq.(\ref{CP asymmetry}) we can get the expression of $A_{CP}(B^-\rightarrow K^- f_0(500))$. Substituting Eqs. (\ref{Mbfk}) and (\ref{e2}) into Eq. (\ref{e1}), one can obtain the branching fraction of $B^-\rightarrow K^- f_0(500)$. Both of them are functions of $X (\rho_S,\phi_S)$.

 \section{Numerical results}
The expressions for $A_{CP}({B^-\rightarrow K^-\pi^+\pi^-})$, $A_{CP}(B^-\rightarrow K^- f_0(500))$ and $\mathcal{B}(B^-\rightarrow K^- f_0(500))$ obtained in the QCD factorization approach depend on many input parameters including CKM matrix elements, effective Wilson coefficients, light-cone distribution amplitudes of mesons, form factors and decay constants. CKM matrix elements can be expressed in the terms of Wolfenstein parameters $A$, $\lambda$, $\rho$ and $\eta$. In our work, we take values $A=0.811^{+0.023}_{-0.024}$, $\lambda=0.225\pm0.00061$, $\bar{\rho}=0.117\pm0.021$, and $\bar{\eta}=0.353\pm0.013$ with $\bar{\rho}=\rho(1-\frac{\lambda^2}{2}), \bar{\eta}=\eta(1-\frac{\lambda^2}{2})$ \cite{Agashe:2014kda}. The effective Wilson coefficients used in our calculations are taken from Ref. \cite{Wang:2014hba}:
\begin{equation}\label{C}
\begin{split}
&C'_1=-0.3125, \quad C'_2=-1.1502, \\
&C'_3=2.120\times10^{-2}+5.174\times10^{-3}i,\quad C'_4=-4.869\times10^{-2}-1.552\times10^{-2}i, \\
&C'_5=1.420\times10^{-2}+5.174\times10^{-3}i,\quad C'_6=-5.792\times10^{-2}-1.552\times10^{-2}i, \\
&C'_7=-8.340\times10^{-5}-9.938\times10^{-5}i,\quad C'_8=3.839\times10^{-4}, \\
&C'_9=-1.017\times10^{-2}-9.938\times10^{-5}i,\quad C'_{10}=1.959\times10^{-3}. \\
\end{split}
\end{equation}

The twist-2 light-cone distribution amplitudes (LCDA) for the pseudoscalar $(P)$ and vector $(V)$ mesons are
\begin{equation}
\Phi_{P,V}(x,\mu)=6x(x-1)\left[1+\sum_{m=1}^\infty \alpha_m^{P,V}(\mu)C_m^{3/2}(2x-1)\right],
 \end{equation}
and twist-3 ones are
\begin{equation}
\Phi_p(x)=1,\quad \Phi_\sigma(x)=6x(x-1),
 \end{equation}
\begin{equation}
\Phi_\upsilon(x)=3\left[2x-1+\sum_{m=1}^\infty \alpha_{m,\perp}^V(\mu)P_{m+1}(2x-1)\right],
 \end{equation}
where $C_m^{3/2}$ and $P_m$ are the Gegenbauer and Legendre polynomials, respectively, $\alpha_m^{P,V}(\mu)$ and $\alpha_{m,\perp}^V(\mu)$ are Gegenbauer moments which depend on the scale $\mu$. The Gegenbauer moments of $K$ and $\rho$ are $\alpha_1^K=0.06\pm0.03$, $\alpha_2^K=0.25\pm0.15$, and $\alpha_1^\rho=0$, $\alpha_2^\rho=0.14\pm0.06$, $\alpha_{1,\perp}^{\rho}=0$, and $\alpha_{2,\perp}^{\rho}=0.15\pm0.07$ \cite{Cheng:2009cn}, respectively, at the scale $\mu=1\ \mathrm{GeV}$.

In general, the twist-2 LCDA of a scalar meson, $\Phi_S$, has the following form \cite{Cheng:2005nb} :
\begin{equation}
\Phi_S(x,\mu)=\bar{f}_S6x(x-1)\sum_{m=1}^\infty B_m(\mu)C_m^{3/2}(2x-1),
 \end{equation}

where$\bar{f}_S$ are the decay constants of the scalar meson $S$, $B_m$ are Gegenbauer moments. Based on the QCD sum rule methods \cite{Govaerts:1986ua,Cheng:2005ye}, we can derive the  decay constants  $\bar{f}^q_{f_0(500)}$ $(q=u,s)$ with the $f_0(980)$-$f_0(500)$ mixing angle $\theta=17^0$ \cite{Cheng:2013fba}: $\bar{f}^u_{f_0(500)}=(0.4829\pm0.076)\ \mathrm{GeV}$ and $\bar{f}^s_{f_0(500)}=(-0.21\pm0.093)\ \mathrm{GeV}$, and Gegenbauer moments: $B_1^u=-0.42\pm0.02$, $B_3^u=-0.58\pm0.19$, $B_1^s=-0.35\pm0.003$, and $B_3^s=-0.43\pm1.26$ at the scale $\mu=1 \ \mathrm{GeV}$. 

 As for the twist-3 distribution amplitudes, we use \cite{Cheng:2005nb}
\begin{equation}
\Phi_S^s(x)=\bar{f}_S,\quad \Phi_S^\sigma(x)=\bar{f}_S6x(x-1).
 \end{equation}

For the form factors of mesons, we neglect corrections quadratic in the light meson masses and we adopt the values at $q^2=0$ in Ref. \cite{Beneke:2003zv} (At this kinematic point, the form factors $F_+$ and $F_0$ coincide.), $A_0^{B\rightarrow\rho}(0)=0.303\pm0.029$, $F_+^{B\rightarrow K}(0)=F_0^{B\rightarrow K}(0)=0.35\pm0.04$ \cite{Cheng:2009cn}. Since most of the models indicate that the $B$ meson to a light meson form factor at zero recoil $q^2$ lies around $0.3$, we simply set $F_0^{B\rightarrow f}(m_K^2)\approx F_0^{B\rightarrow f}(0)=0.3$ and assign its uncertainty to be $\delta F^{BS}(0)=\pm0.03$ \cite{Cheng:2005nb}. The decay constants used in our calculations are $f_K=0.156\pm0.7 \mathrm{GeV}$ \cite{Agashe:2014kda}, $f_\rho=0.216\pm0.003 \mathrm{GeV}$, and $f_B=0.21\pm0.02 \mathrm{GeV}$ \cite{Cheng:2009cn}.

 A general fit of $\rho$ and $\phi$ to the $B\rightarrow VP$ and $B\rightarrow PV$ data indicates $X^{PV}\neq X^{VP}$, i.e. $\rho^{PV}\approx0.87$, $\rho^{VP}\approx1.07$, $\phi^{VP}\approx-30^0$ and $\phi^{PV}\approx-70^0$, we shall assign an error of $\pm0.1$ to $\rho^{PV(VP)}$ and $\pm20^0$ to $\phi^{PV(VP)}$ for the estimation of theoretical uncertainties  \cite{Cheng:2009cn}. On the other hand, for $B\rightarrow PS$ and $B\rightarrow SP$ decays, there is little experimental data so the values of $\rho_S$ and $\phi_S$ are not determined very well,  to make an estimation about $A_{CP}(B^-\rightarrow K^- f_0(500))$ and $\mathcal{B}(B^-\rightarrow K^-f_0(500))$, we adopt $X^{PS}=X^{SP}=(1+\rho_S e^{i\phi_S})\ln\frac{m_B}{\Lambda_h}$.

With all the above considerations, we only have three free parameters, which are the relative strong phase $\delta$, and the divergence parameters $\rho_S$ and $\phi_S$ for $A_\emph{CP}(B^-\rightarrow K^-\rho^0(770)(f_0(500))\rightarrow K^-\pi^+\pi^-)$. By fitting the theoretical result to the experimental data $A_\emph{CP}(B^-\rightarrow K^-\pi^+\pi^-)=0.678\pm0.078\pm0.0323\pm0.007$ in the region $m_{K^-\pi^+}^2<15$ $\mathrm{GeV}^2$ and $0.08<m_{\pi^+\pi^-}^2<0.66$ $\mathrm{GeV}^2$, in the range $\delta\in[0,2\pi]$, $\phi_S\in[0,2\pi]$, $\rho_S\in[0,8]$ \cite{Ciuchini:2002uv} and varying each of these three parameters by 0.01 each time, i.e.  $\Delta\delta=0.01$, $\Delta\rho_S=0.01$ and $\Delta\phi_S=0.01$, it is found that there exist ranges of parameters $\delta$, $\rho_S$ and $\phi_S$ which satisfy the above experimental data. The allowed ranges are $\delta\in[0, 1.745]\cup[3.578, 6.266]$, $\rho_S\in[2.790, 5.290]$ and $\phi_S \in [1.518, 5.183]$. Therefore, the interference of $\rho^0(770)$ and $f_0(500)$ can indeed induce the data for the localized $CP$ asymmetry in the $B^-\rightarrow K^-\pi^+\pi^-$ decays. It is noted that the values of $\rho_S\in[2.790, 5.290]$ are relative larger compared with the previously conservative choice of $\rho\leq1$ \cite{Beneke:2001ev,Beneke:2003zv}. Because the QCDF itself cannot give information about parameters $\rho$ and $\phi$, there is no reason to restrict $\rho$ to the range $\rho\leq1$ \cite{Cheng:2009cn,Chang:2014rla,Sun:2014tfa,Zhu:2011mm}, thus larger values of $\rho_S$ are acceptable to deal with the divergence problems for $B\rightarrow SP(PS)$ decays. In this region of $\rho_S$, one can see that the weak annihilation and the hard spectator scattering processes can make large contributions to $B^-\rightarrow K^- f_0(500)$ decays.

\begin{figure}[ht]
\begin{minipage}{0.4\linewidth}
\centerline{\includegraphics[width=1\textwidth]{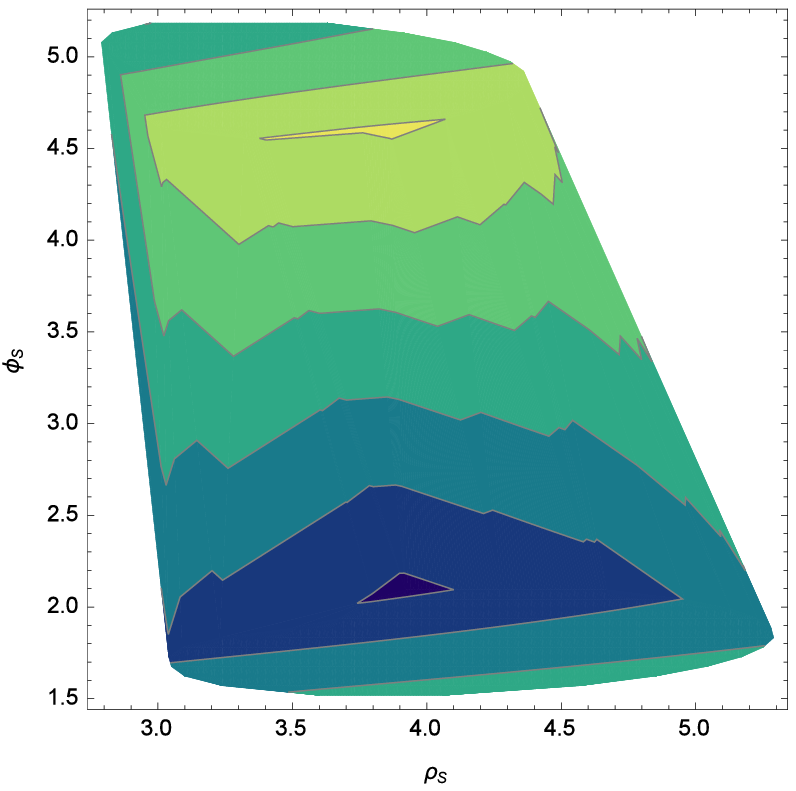}}
\centerline{}
\end{minipage}
\qquad
\begin{minipage}{0.07\linewidth}
\centerline{\includegraphics[width=1\textwidth]{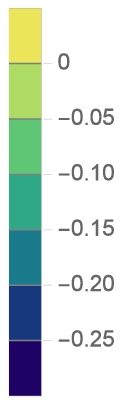}}
\centerline{}
\end{minipage}
\caption{Numerical results of $A_{CP}(B^{\pm}\rightarrow f_0(500)K^{\pm})$ as functions of $\rho_S$ and $\phi_S$.}
\label{p1}
\end{figure}

\begin{figure}[ht]
\begin{minipage}{0.4\linewidth}
\centerline{\includegraphics[width=1\textwidth]{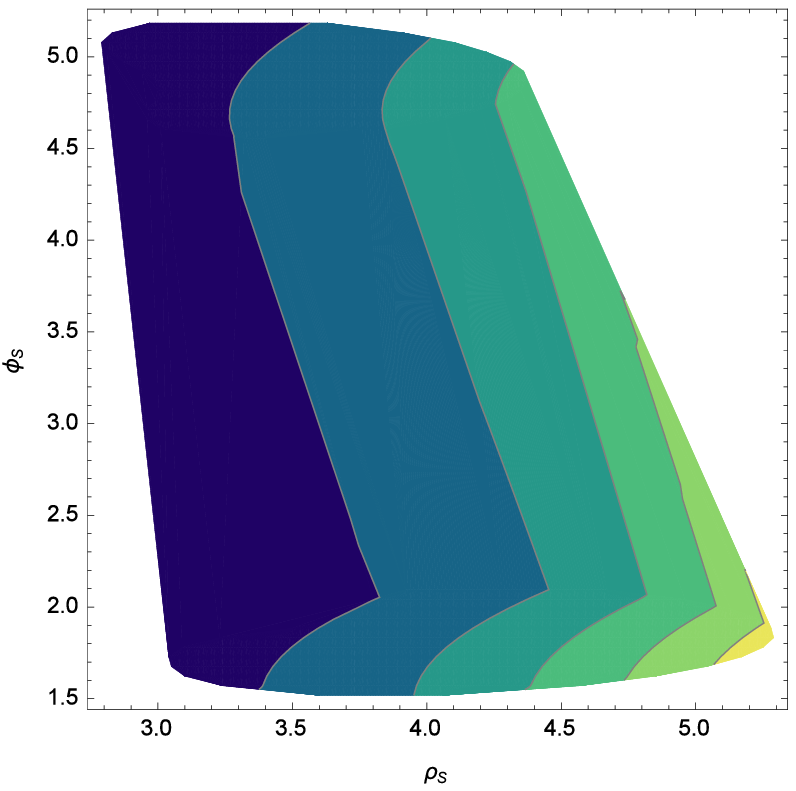}}
\centerline{}
\end{minipage}
\qquad
\begin{minipage}{0.06\linewidth}
\centerline{\includegraphics[width=1\textwidth]{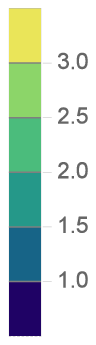}}
\centerline{}
\end{minipage}
\caption{Numerical results of $\mathcal{B}(B^-\rightarrow f_0(500)K^-)$ $(\times10^5)$ as functions of $\rho_S$ and $\phi_S$.}
\label{p2}
\end{figure}

In the obtained allowed ranges for $\rho_S$ and $\phi_S$, i.e. $\rho_S\in[2.790, 5.290]$ and $\phi_S \in [1.518, 5.183]$, we calculate the $CP$ asymmetry parameter and the branching fraction for $B^-\rightarrow K^-f_0(500)$ decay modes using Eqs. (\ref{CP asymmetry}), (\ref{Mbfk}), (\ref{e1}) and (\ref{e2}). The results are plotted in Figs. \ref{p1} and \ref{p2} as functions of $\rho_S$ and $\phi_S$. From these two figures and our calculated data, we obtain the predictions $A_{CP}(B^-\rightarrow K^-f_0(500))\in [-0.259, 0.006]$ and $\mathcal{B}(B^-\rightarrow K^-f_0(500))\in [0.585, 3.230]\times10^{-5}$ when $\rho_S$ and $\phi_S$ vary in their allowed ranges.

\section{SUMMARY}
In this work, within the QCD factorization approach, we study the localized $CP$ violation in $B^-\rightarrow K^-\pi^+\pi^-$ decays in the region $m_{K^-\pi^+}^2<15$ $\mathrm{GeV}^2$ and $0.08<m_{\pi^+\pi^-}^2<0.66$ $\mathrm{GeV}^2$ by including the interference of $\rho^0(770)$ and $f_0(500)$. By fitting the experimental data of $A_{CP}(B^-\rightarrow K^-\pi^+\pi^-)$ in this region, we find that such localized $CP$ asymmetry can be indeed induced by the interference of $\rho^0(770)$ and $f_0(500)$ when $\delta\in[0, 1.745]\cup[3.578, 6.266]$, $\phi_S \in [1.518, 5.183]$ and $\rho_S\in[2.790, 5.290]$. The large values of $\rho_S$ indicate that the weak annihilation and the hard spectator scattering processes can make large contributions and we should take more efforts to investigate these contributions in $B$ nonleptonic weak decays. With the obtained allowed ranges for $\rho_S$ and $\phi_S$, we predict the $CP$ asymmetry parameter and the branching fraction for $B^-\rightarrow K^-f_0(500)$ decay modes. We find  $A_{CP}(B^-\rightarrow K^-f_0(500))\in [-0.259, 0.006]$ and $\mathcal{B}(B^-\rightarrow K^-f_0(500))\in [0.585, 3.230]\times10^{-5}$ in the allowed ranges of  $\phi_S$ and $\rho_S$. These predictions can hopefully be tested in future experiments. In our analysis, the uncertainties coming from the CKM matrix elements, form factors, decay constants, $s$ quark masses and Gegenbauer moments are all considered.

\acknowledgments
 One of the authors (J.-J. Qi) is very grateful to thank Professor Hsiang-nan Li for valuable discussions. This work was supported by National Natural Science Foundation of China (Projects No. 11575023, No. 11775024, No. 11705081 and No. 11605150).

\end{document}